\documentclass{sf2a-conf2012}
\usepackage{graphicx}
\usepackage[]{natbib}  
\usepackage{amsmath,amssymb,amsfonts,amsthm}
\usepackage{epstopdf}

\usepackage{color}
\definecolor{darkgreen}{RGB}{0,142,128}
\usepackage{hyperref}
\hypersetup{colorlinks,citecolor=blue,linkcolor=blue}

\usepackage{multirow}
\usepackage{subfigure}
\def\BibTeX{{\rm B\kern-.05em{\sc i\kern-.025em b}\kern-.08em
    T\kern-.1667em\lower.7ex\hbox{E}\kern-.125emX}}
\bibpunct{(}{)}{;}{a}{}{,}  %%%%%%%%%%%%%  A&A bibliography style
\begin{document}

\TitreGlobal{SF2A 2012}

%%-----------------------------------------------------------------
%%      the top matter
%%

\title{On close-in magnetized star-planet interactions}

\runningtitle{On close-in magnetized star-planet intractions}

\author{A. Strugarek}\address{Laboratoire AIM Paris-Saclay, CEA/Irfu Universit\'e Paris-Diderot CNRS/INSU, F-91191 Gif-sur-Yvette.}

\author{A. S. Brun$^1$}
\author{S. Matt$^1$}

%% Keep this line, even if the page will be settled afterwards.
\setcounter{page}{237}

%%-----------------------------------------------------------------

\maketitle

%%-----------------------------------------------------------------
%%        The abstract
%% 
%%  Warning!  within the abstract:
%%  - do not use macros. 
%%  - do not use commands like: \cite, \citet, \citep ... etc.

\begin{abstract}
We present 2D magnetohydrodynamic simulations performed with the PLUTO
code to model magnetized star-planet interactions. We study two
simple scenarios of magnetized star-planet interactions: the
\textit{unipolar} and \textit{dipolar} interactions. Despite
the simplified hypotheses we consider in the model, the
qualitative behavior of the interactions is very well
recovered. These encouraging results promote further developments of
the model to obtain predictions on the effect and the physical
manifestation of magnetized star--close-in planet interactions.
\end{abstract}

%% Insert the keywords (to appear in the ADS indexing)
%% Keywords must be separated by a comma

%%-----------------------------------------------------------------

\section{Introduction}
\label{sec:introduction}

More than $830$ exo-planets have been presently
discovered\footnote{\url{http://exoplanet.eu}}.
The interactions between stars and their orbiting planets can be
distinguished between \textit{distant} gravitational (orbital motions,
tides) and \textit{direct}
(hydro-)magnetic (stellar wind, radiation) interplays. Both interactions are
likely to play a major role in determining habitability zones and in
understanding planet dynamics. In addition, close-in giant planets
may also impact the rotation and magnetic properties of their host
stars \citep{Donati:2008hw,Bolmont:2012go}. Finally, the magnetized interactions can yield
enhanced localized emissions in the chromospheres of their
host stars \citep{Shkolnik:2005bz}. For these reasons, a better
characterization of star-planet 
interactions (SPIs) would be highly valuable
\citep{Cuntz:2000ef,Ip:2004ba,Lanza:2009fp}. In
this paper, we focus on 
basic mechanisms that underly the \textit{direct} magnetized SPIs.

Magnetized SPIs can be separated into two classes: the so-called
\textit{unipolar} (magnetized wind, unmagnetized planet) and
\textit{dipolar} (magnetized wind and planet) interactions
\citep{Zarka:2007fo}. They were initially studied in the context of
satellites orbiting in the magnetosphere of giant planets
\citep{Kivelson:2004vf}. The space plasma in the upper magnetospheres
of planets and in stellar winds is characterized by a Knudsen number
(mean free path over characteristic size of the
system) much greater than unity. Hence, the fluid approximation does
not hold because there is \textit{a priori} no
reason to consider that the plasma is locally thermally equilibrated:
a kinetic modeling should be used to accurately represent it
\citep{Marsch:2006vp}. Because of limited
computation resources, the large scales involved in
magnetized SPIs prevent us to use a global kinetic modeling. 
Magnetohydrodynamic (MHD) models (which are less expensive to
simulate) have been therefore widely used instead. Such models are able to
recover the global properties of 
stellar winds \citep{Goldstein:1995ei} and have been used
either by fitting the equation of state to recover the exact solar
wind \citep{Wang:1990kl,Arge:2000gd}, or by conducting parametric
studies to derive general scaling laws 
\citep[][and references therein]{Washimi:1993vm,Matt:2012ib}. \citet{Cohen:2011gg}
simulated magnetized SPIs based on the former kind of modeling (using
the so-called \textit{WSA} model). In this paper, we base our
study on the latter modeling approach, which will allow us to derive
robust scaling laws for magnetized SPIs.

We develop in section \ref{sec:model-magn-star} the method we use to
study magnetized SPIs. Then, we apply our setup to the two basic
cases of the unipolar and diploar interactions in section
\ref{sec:class-inter}. We validate the modeling choices we made and
are able to predict the action of a 
close-in planet on the stellar surface flows. Finally, we give the
perspectives of this preliminary work in section \ref{sec:perspectives}.

\section{Modeling magnetized star-planet interactions}
\label{sec:model-magn-star}

The magnetized SPIs consists of the interaction between the magnetized wind of the
host star and the magnetized or unmagnetized planet. Any modeling tackling
these interactions have to treat plasmas associated with both the wind and the planet.

\subsection{Wind modeling}
\label{sec:wind-modeling}

Following numerous previous studies
\citep[\textit{e.g.},][]{Ustyugova:1999ig}, we use standard MHD wind
theory 
that characterizes magnetized steady-state flows anchored at the
surface of a rotating star. The exact wind driving mechanism is still
debated today, its details should not matter for the purpose of this
paper. Hence we make the assumption that it is driven by 
the thermal pressure of the coronal plasma (which is a common basic
assumption, \textit{e.g.}, in the case of the solar wind). 

We use the PLUTO code \citep{Mignone:2007iw} to calculate steady-state
winds using the ideal compressible MHD equations (written here in
their primitive formulation for simplicity)
\begin{eqnarray}
  \label{eq:mass_consrv_pluto}
  \partial_t \rho + \boldsymbol{\nabla}\cdot(\rho \mathbf{v}) &=& 0 \, \\
  \label{eq:mom_consrv_pluto}
  \partial_t\mathbf{v} +
  \mathbf{v}\cdot\boldsymbol{\nabla}\mathbf{v}+\frac{1}{\rho}\boldsymbol{\nabla} P
  +\frac{1}{\rho}\mathbf{B}\times\boldsymbol{\nabla}\times\mathbf{B}
  &=& \mathbf{g} \, ,
  \\
  \label{eq:ener_consrv_pluto}
  \partial_t P +\mathbf{v}\cdot\boldsymbol{\nabla} P + \rho
  c_s^2\boldsymbol{\nabla}\cdot\mathbf{v} &=& 0 \, ,\\
  \label{eq:induction_pluto}
  \partial_t \mathbf{B} - \boldsymbol{\nabla}\left(\mathbf{v}\times\mathbf{B}\right)
  &=& 0 \, ,
\end{eqnarray}
where $\rho$ is the density, $\mathbf{v}$ the velocity, $P$ the gas
pressure, $\mathbf{B}$ the magnetic field, $\mathbf{g}$ the
gravitational acceleration, $\partial_t$ the derivative with respect
to time and $c_s=\sqrt{\gamma\,P/\rho}$ the sound
speed ($\gamma$ is the polytropic index of the plasma). We use an
ideal gas equation of state. 

We use the following options in PLUTO to run our simulations. A minmod
limiter on all the variables and a \textit{hll} 
(Harten, Lax, Van Leer) solver to compute the intercell fluxes. A
second order Runge-Kutta scheme is used for the time
evolution. The solenoidality of the magnetic field
($\boldsymbol{\nabla}\cdot\mathbf{B}=0$) is ensured with a
\textit{constrained transport} (CT) method \citep[\textit{e.g.},][]{Gardiner:2005ky}.

We initialize our simulation with the spherically symmetric
hydrodynamic Parker solution \citep{Parker:1958dn}. We add a dipolar
magnetic field characterized by the ratio $v_A/v_{\rm esc} = 0.32$ at
the surface of the star ($v_A=B/\sqrt{4\pi\rho}$ is the Alfv\`en speed and $v_{\rm
  esc}=\sqrt{2GM_\star/R_\star}$ is the escape speed). We developed
special boundary conditions \citep{Matt:2008bj,Zanni:2009kc} that
ensure the conservation of the five quantities theoretically conserved along field
lines that were identified by
\citet{Lovelace:1986kd} and \citet{Keppens:2000ea}. The
parameters and characteristics of the simulated wind are given in
table \ref{tab:tab1}. The wind exhibits a large dead zone (closed
field lines and very slow motions region) inside its alfvenic
surface that extends up to $r\sim 7\, R_\star$ at the equator. We
develop hereafter the methodology we use to simulate an orbiting planet.

\begin{table}[ht!]
  \centering
  \begin{tabular}{ccccc}
    \hline
    \hline
    \multirow{2}{*}{Parameters} & $\gamma$ & $c_{\rm s}/v_{\rm esc}$ & $v_{\rm rot}/v_{\rm esc}$
    & $v_{\rm A}/v_{\rm esc}$ \\
    & 1.05 & 0.2220 & 0.00303 & 0.3183 \\
    \hline
    \multirow{2}{*}{Characteristics} & $\dot m$ & $\Psi_{\rm o}/\Psi_\star$ & $\dot j$
    & $v_{\rm p}/v_{\rm esc}$ (15$r_*$, $45^\circ$) \\
    & 6.48 $10^{-4}$ & 0.251 & 7.12 $10^{-4}$ &
    0.250 \\
    \hline
  \end{tabular}
  \caption{Parameters and deduced characteristics of the simulated
    stellar wind. The parameter $v_{\rm rot}/v_{\rm esc}$ sets the
    rotation rate of the star. The mass loss rate is normalized to
    $\rho_\star v_{\rm esc} R_\star^2$ and the angular momentum loss
    rate to $\rho_\star v_{\rm esc}^2 R_\star^3$. The ratio between
    opened and closed field lines 
    is given by $\Psi_{\rm o}/\Psi_\star$ ($1-\Psi_{\rm o}/\Psi_\star$
    gives the
    size of the dead zone, with $\Psi_{\rm o}\equiv \oint
    \mathbf{B}\cdot{\rm d}\mathbf{A}$). Finally, we give the typical poloidal
    velocity of the wind at $(r=15\,R_\star,\theta=\,45^\circ)$.}
  \label{tab:tab1}
\end{table}

\subsection{Planet modeling}
\label{sec:planet-modeling}

We introduce a planet as a boundary condition inside the computational
domain. We choose to study only close-in planets in this paper,
\textit{i.e.} planets which are \textit{inside} the dead-zone of the
stellar wind. Hence, we introduce a very close planet at $r_{\rm
  orb}=2.5\, R_\star$. The type of
interaction between the two bodies is determined by the
topology of the planetary magnetic field. We choose a
heavy Jupiter-like planet such that $r_P=0.1\, R_\star$
($R_J\sim0.1\,R_\odot$) and $m_p = 0.01\,M_\star$
($M_J=0.001\,M_\odot$). We design a stretched grid such that the
typical resolution is of the order of $r_P/32$ at the planet surface
and of $R_\star/64$ at the stellar surface. We ensured the numerical convergence
of our results when increasing the resolution by a factor two.

The initialization of the planet in the steady-state wind creates a
transient evolution that is rapidly forgotten. Because we use an
idealized axisymmetric configuration (the so-called 2.5D
approximation, \textit{i.e.}, we study the 3D fields only on an axisymmetric
poloidal plane), the planet we simulate has the shape of a torus
circling the star, rather than a sphere. As a consequence, the orbital motion does not introduce any
time-variability in the orbital direction and a new steady-state
can be obtained. Even if this situation is far from reality, it
constitutes a first step towards the realistic modeling of magnetized SPIs
(see perspectives in section \ref{sec:perspectives}).

\section{Basic interactions}
\label{sec:class-inter}

As mentioned in the introduction, the magnetized SPIs can be decomposed into
\textit{unipolar} and \textit{dipolar} interactions \citep{Zarka:2007fo}. We successively
simulate the two situations in the following, which
are very well recovered by our model.

% \subsection{Unipolar interaction}
% \label{sec:unipolar-interaction}

\begin{figure}[ht!]
 \centering
 \subfigure[]{
   \includegraphics[width=0.4\textwidth,clip]{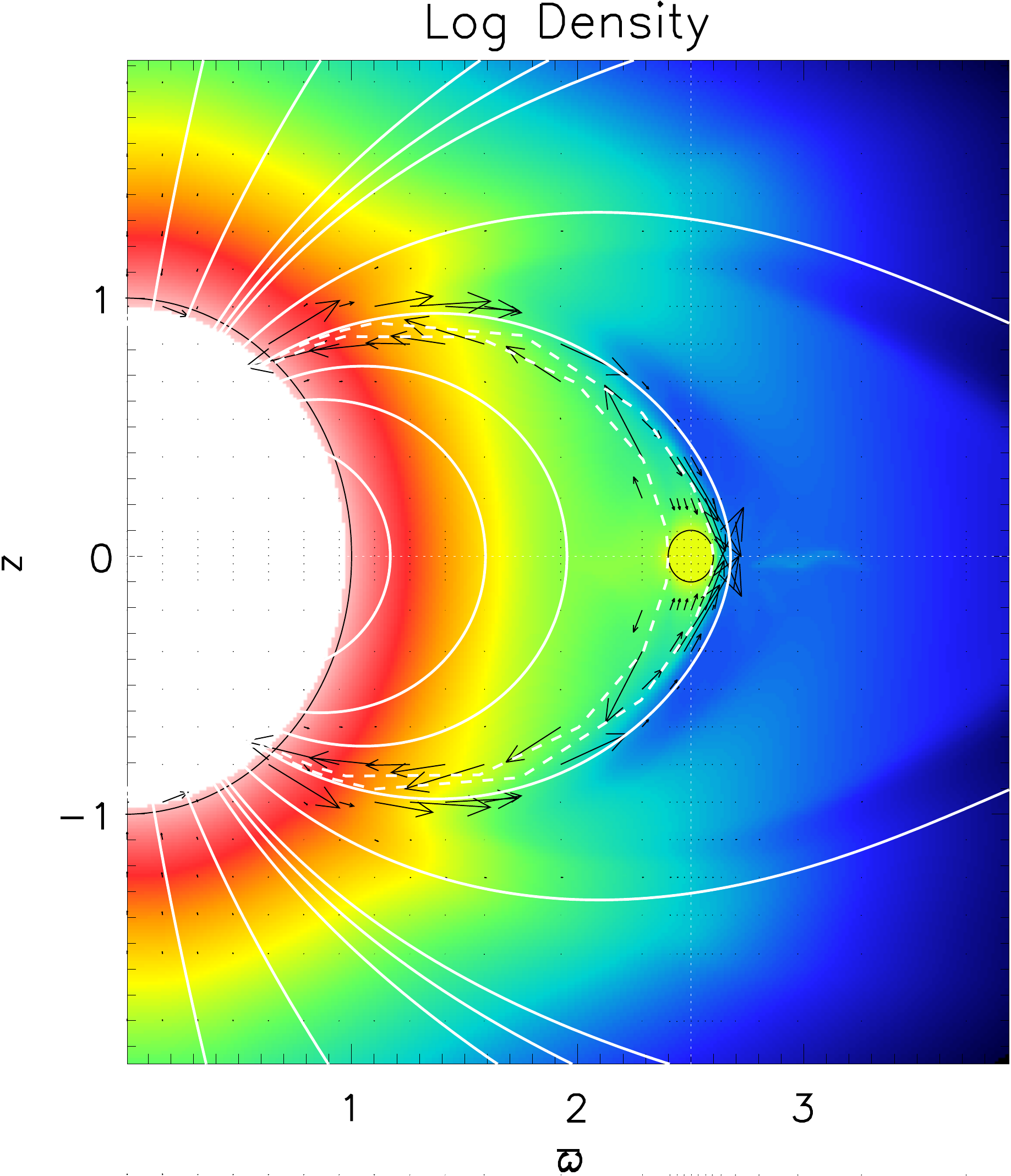}      
  \label{fig:unipolar}
 }
 \subfigure[]{
 \includegraphics[width=0.4\textwidth,clip]{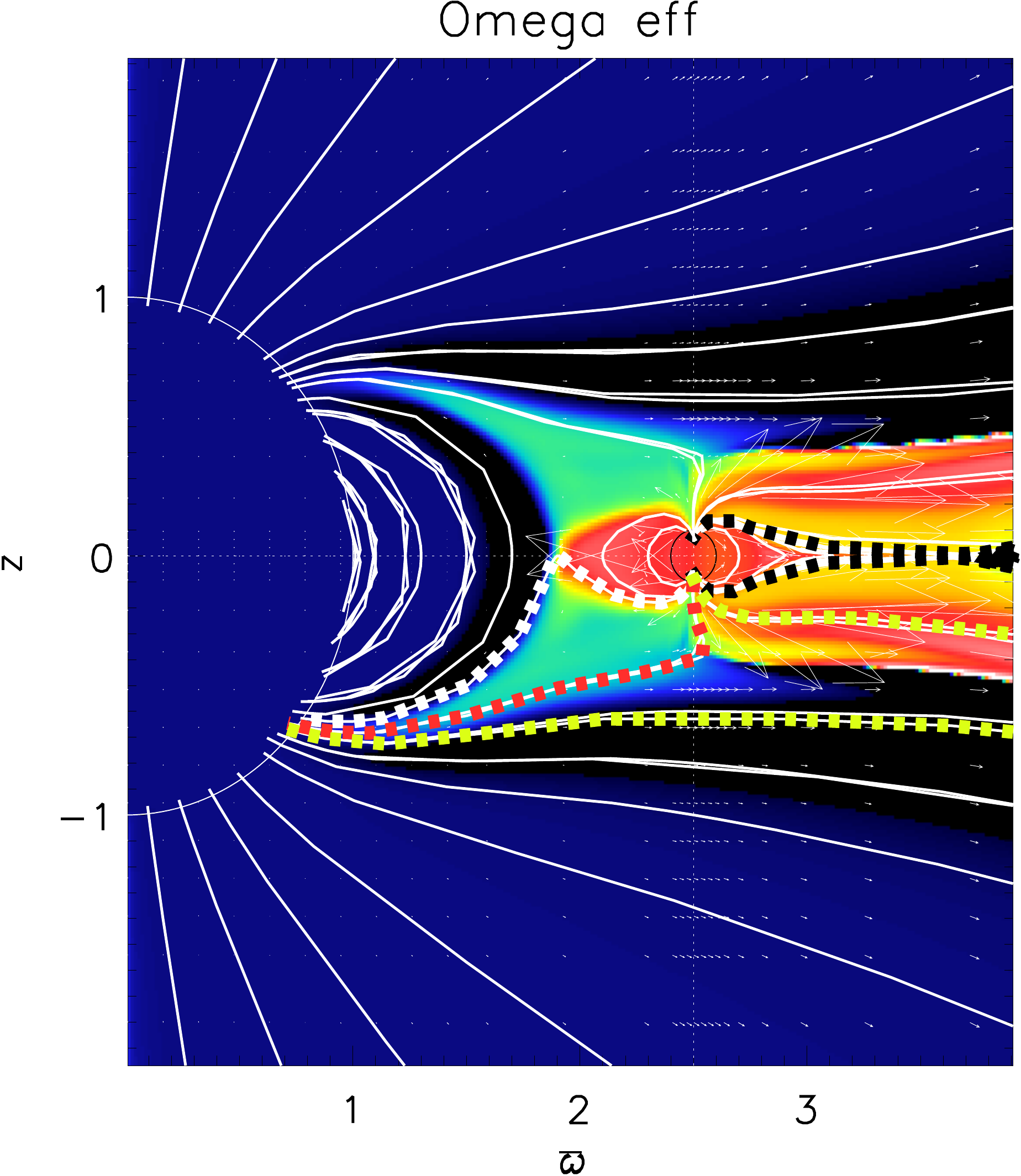}      
  \label{fig:dipolar}
}
  \caption{\textit{(a)} Simulation of the unipolar interaction. The
    color map represents the logarithm of density (blue/black is low
    density, red/white high density) and the black arrows are the current
    density $\mathbf{J}$. \textit{(b)} Simulation of the dipolar
    interaction. The colormap represent the effective rotation rate
    $\Omega_{\rm eff} =
    \frac{1}{\varpi}\left(v_\varphi-\frac{\mathbf{v}_p\cdot\mathbf{B}_p}{B_p^2}B_\varphi
    \right)$, where $(\varpi,\varphi,z)$ are the cylindrical
    coordinates and the subscript $p$ denotes the poloidal
    $(\varpi,z)$ component of the field. The rotation rate of the star
    is blue. The white arrows are the
    velocity field.}
  \label{fig:fig1}
\end{figure}

\paragraph{-- Unipolar interaction --}
\label{sec:unipolar-interaction-1}
We introduce a unmagnetized rotating planet in the dead zone of the
simulated wind (section \ref{sec:wind-modeling}). This system is
equivalent to the well known interaction of Io in the magnetosphere of
Jupiter. The unmagnetized planet drags the poloidal magnetic field
lines and current sheets establish along the poloidal field lines connecting
the planet to the host star \citep{Goldreich:1969kf}. We indeed
observe a current loop in figure \ref{fig:unipolar} that connect the
planet and the star together (the black arrows represent the current
density $\mathbf{J}=\boldsymbol{\nabla}\times\mathbf{B}$). An azimuthal
component of the magnetic field ($B_\varphi$) is also naturally created
through an effective 'Omega'-effect generated by the orbital motion of
the planet. A steady state situation is achieved when the numerical
diffusion of the magnetic field in the azimuthal direction is balanced
by the continuous twisting action of the differential rotation between
the star and planet orbit. A steady
state return flow from the planet to the star is then associated with
the magnetic 
flux-tube. The magnetic connection between the star and the planet
implies the existence of a torque between the two objects. The planet
orbits at the keplerian velocity (such that its orbital motion
compensates the stellar gravitational pull), hence it rotates much
faster than the stellar surface (\textit{i.e.}, the co-rotation radius
is larger than the orbital radius of the planet). The planet exerts consequently a torque localized
in latitude which is approximately 4 times larger (and of opposite
sign) than the overall torque 
exerted by the stellar wind when no planet is taken into account. The
exact value of the
torque is likely to depend on both \textit{(i)} the fact that we are
considering a 2D setup and \textit{(ii)} the amount of numerical
diffusivity.

% \subsection{Dipolar interaction}
% \label{sec:dipolar-interaction}

\paragraph{-- Dipolar interaction --}
\label{sec:-dipolar-interaction}
We also perform the exact same simulation 
with a \textit{magnetized} planet. We choose a dipolar planetary
magnetic field anti-aligned with the initial stellar dipole (we choose its
original amplitude such that the initial planetary magnetosphere is of
the order of $2\, r_p$). As a
consequence, the closed magnetic field lines of the wind naturally connect at the poles
of the planet, and magnetic reconnection occur at the equator where
the magnetic field lines of the wind and of the planet are
anti-aligned. We also note here that the planetary field only
resembles a dipole in the poloidal plane but was slightly modified to
preserve $\boldsymbol{\nabla}\cdot \mathbf{B}=0$ in an axisymmetric geometry.
 We display in figure \ref{fig:dipolar} the interaction
between the magnetized planet and the stellar wind.

Reconnections of the magnetic field lines occur at the equator and are
labeled by the dashed white line. We see that the field lines connect
together the planet and the stellar surfaces (dashed red
line). Because these field lines are close to the closed-opened field
lines boundary, they tend to be advected by the stellar wind and are
stretched away from the planet (yellow dashed line). Reconnection then
occur again  and the magnetic field lines close in the magneto-tail of
the planetary magnetosphere (black dashed line). This process is very
similar to the basic reconnection mechanism developed to explain the
structure of the magnetosphere of the planets of the solar system
\citep{Gombosi:1998tj}.

We recall here that these numerical experiments are done in the
framework of ideal MHD. Hence, any reconnection occurring in the
simulations is controlled by the effective diffusion introduced by the
numerical techniques we use. In order to quantitatively characterize the
reconnection process we observe, a better control of the ohmic diffusion is
mandatory and will be adressed in future work.

Finally, the magnetic connection between the two objects
is stronger than in the unipolar case and the torque exerted by the
planet on the stellar surface is roughly twenty times larger (and
of opposite magnitude) than the torque exerted by the stellar wind.

\section{Conclusions and perspectives}
\label{sec:perspectives}

In this paper, we demonstrated that ideal MHD simulations in 2D
axisymmetric geometry could well capture the basic magnetized SPIs
involving a close-in giant planet orbiting inside the alfvenic
surface of its host star. We tested both the unipolar (unmagnetized planet) and dipole
(magnetized planet) interactions and showed that the former were
likely to exert a greater torque on the stellar surface. Because the
planet orbital motion and the rotation rate of the star are fixed, the applied torque
does not modify the surface rotation nor the planet orbit. Fixing them is
legitimate here since approximately $10^{14}$ orbits would be required to change the orbital radius
by $0.1\, R_\star$ in the unipolar case (based on the observed torque
in the simulations). This picture may drastically change when varying
the wind and planet parameters.

We established a modeling framework that will allow us to develop a
complete numerical analysis of magnetized SPIs. The obvious next step
consist naturally in simulating the star-planet pairs in 3D in order to
let the interaction develop in the correct geometry. Then, we will be
able to explore the various interaction regimes depending on the
magnetic topologies and time-variability of the stellar and planetary
fields, and on the position of the planetary orbit in the stellar
wind. Finally, we also intend to develop tools to determine
the expected level of emissions resulting from the magnetized
SPIs \citep[\textit{e.g.}, see][]{2012arXiv1204.3843V}. This work will
lead to reliable scaling laws on the effect of 
magnetized SPIs that will be useful to explain and guide
exoplanet observations, but also to test fundamental ideas explaining the physical
processes underlying these interactions.

% Optional acknowledgements
% -------------------------
\begin{acknowledgements}
The authors thank N. Bessolaz, R. Pinto and C. Zanni for very helpful
discussions at the origin of this work.
\end{acknowledgements}

%% The following lines are required when using BibTEX (strongly encouraged!):
\bibliographystyle{aa}  % A&A bibliography style file (aa.bst)
\bibliography{mybib} % your references in file: Yourfile.bib

\end{document}